\documentclass[a4paper,11pt]{article}
\usepackage{pos}
\usepackage{float}
\usepackage{cancel}
\usepackage{titlesec}
\usepackage{braket}
\usepackage{multicol}
\usepackage{hyperref}
\usepackage{subfigure}
\usepackage{graphicx}      

\title{Tensor Charges and their Impact on Physics Beyond the Standard Model}

\author*[a]{R. E. Smail}
\author[b]{R. Horsley}
\author[c]{Y. Nakamura}
\author[d]{H. Perlt}
\author[e]{D. Pleiter}
\author[f]{P. E. L. Rakow}
\author[g]{G. Schierholz}
\author[h]{H. St$\ddot{\text{u}}$ben}
\author[a]{R. D. Young}
\author[a]{J. M. Zanotti}

\affiliation[a]{CSSM, Department of Physics, University of Adelaide, Adelaide SA 5005, Australia}
\affiliation[b]{School of Physics and Astronomy, University of Edinburgh, Edinburgh EH9 3FD, UK}
\affiliation[c]{RIKEN Center for Computational Science, Kobe, Hyogo 650-0047, Japan}
\affiliation[d]{Institut f$\ddot{\text{u}}$r Theoretische Physik, Universit$\ddot{\text{a}}$t Leipzig, 04103 Leipzig, Germany}
\affiliation[e]{PDC Center for High Performance Computing, KTH Royal Institute of Technology, SE-100 44 Stockholm, Sweden}
\affiliation[f]{Theoretical Physics Division, Department of Mathematical Sciences, University of Liverpool, Liverpool L69 3BX, UK}
\affiliation[g]{Deutsches Elektronen-Synchrotron DESY, Notkestr. 85, 22607 Hamburg, Germany}
\affiliation[h]{Universit$\ddot{\text{a}}$t Hamburg, Regionales Rechenzentrum, 20146 Hamburg, Germany}

\emailAdd{rose.smail@adelaide.edu.au}

\abstract{The nucleon tensor charge, g$_T$, is an important quantity in the search for beyond the Standard Model tensor interactions in neutron and nuclear $\beta$-decays as well as the contribution of the quark electric dipole moment (EDM) to the neutron EDM. We present results from the QCDSF/UKQCD/CSSM collaboration for the tensor charge, $g_T$, using lattice QCD methods and the Feynman-Hellmann theorem. We use a flavour symmetry breaking method to systematically approach the physical quark mass using ensembles that span three lattice spacings.}

\FullConference{%
 The 38th International Symposium on Lattice Field Theory, LATTICE2021
  26th-30th July, 2021
  Zoom/Gather@Massachusetts Institute of Technology
}

\begin{document}
\maketitle

\section{Introduction}
Historically nuclear and neutron beta decays have played an important role in determining the vector-axial (V-A) structure of weak interactions and in shaping the Standard Model (SM). However, more recently neutron and nuclear $\beta$-decays can be used to probe the existence of Beyond the Standard Model (BSM) tensor and scalar interactions. Many experiments are underway worldwide with the aim to improve the precision of measurements of neutron decay observables, two being the neutrino asymmetry $B$ \cite{57030350020}, and the Fierz interference term $b$ \cite{POCANIC2009211,article1}. The parameter $b$ has linear sensitivity to BSM physics through:


\noindent\begin{minipage}{.45\linewidth}
\begin{equation}
\begin{aligned}
b^{\text{BSM}}=&\frac{2}{1+3\lambda^2}\Big[g_S\epsilon_S-12\lambda g_T\epsilon_T\Big]\\
\approx&0.34g_S\epsilon_S-5.22g_T\epsilon_T,
\label{eqn:b}
\end{aligned}
\end{equation}
\end{minipage}%
\begin{minipage}{.54\linewidth}
\begin{equation}
\begin{aligned}
b^{\text{BSM}}_v=&\frac{2}{1+3\lambda^2}\Big[g_S\epsilon_S\lambda-4\lambda g_T\epsilon_T(1+2\lambda)\Big]\\
\approx&0.44g_S\epsilon_S-4.85g_T\epsilon_T,
\label{eqn:bv}
\end{aligned}
\end{equation}
\end{minipage}\\
\newline
where $\epsilon_T$ and $\epsilon_S$ are the new-physics effective couplings, $g_T$ and $g_S$ are the tensor and scalar nucleon isovector charges and $\lambda=g_A/g_V$ \cite{Bhattacharya:2011qm}. Here $b^{\text{BSM}}_v$ is a correction term to the correlation coefficient $B$. Data taken at the Large Hadron Collider (LHC) is currently looking at probing scalar and tensor interactions at the $\lesssim10^{-3}$ level~\cite{Bhattacharya:2011qm}. However to fully assess the discovery potential of experiments at the $10^{-3}$ level it is crucial to identify existing constraints on new scalar and tensor operators.

The quark tensor charges are important quantities when analysing the neutron electric dipole moment (EDM). The neutron EDM is sensitive to CP violation and hence is an excellent probe in the search for physics beyond the Standard Model. CP violating interactions contribute largely to the quark EDM. The dependence of the neutron EDM on the quark EDM, $d_q$, is related to the quark tensor charges, $\delta q$, by \cite{Ellis:1996dg,Bhattacharya:2012bf,PhysRevD.91.074004}:
\begin{equation}
\begin{aligned}
d_n=d_u \delta d+d_d \delta u+d_s \delta s.
\label{eqn:dn}
\end{aligned}
\end{equation}
Here $d_u,~d_d,~d_s,$ are the new effective couplings which contain new CP violating interactions at the TeV scale. The current experimental data gives an upper limit on the neutron EDM of $|d_n| < 2.9 \times10^{-26}e$ cm~\cite{PhysRevLett.97.131801}. In calculating the tensor charges and knowing a bound on $d_n$, we are able to constrain the couplings, $d_q$, and hence BSM theories.\\
QCDSF/UKQCD/CSSM collaborations have an ongoing program investigating various hadronic properties using the Feynman-Hellmann theorem \cite{FH1, FH2, FH3, FH5, Horsley:2012pz, Chambers:2017dov, Can:2020sxc, Hannaford-Gunn:2021mrl}. Here we extend this work to a dedicated study of the nucleon scalar and tensor charges. We discuss a flavour symmetry breaking method to systematically approach the physical quark mass. Finally, we look at the potential impact of our results on measurements of the Fierz interference term.
\section{Simulation Details}
 \begin{figure}[h]
\center
  \includegraphics[scale=0.4]{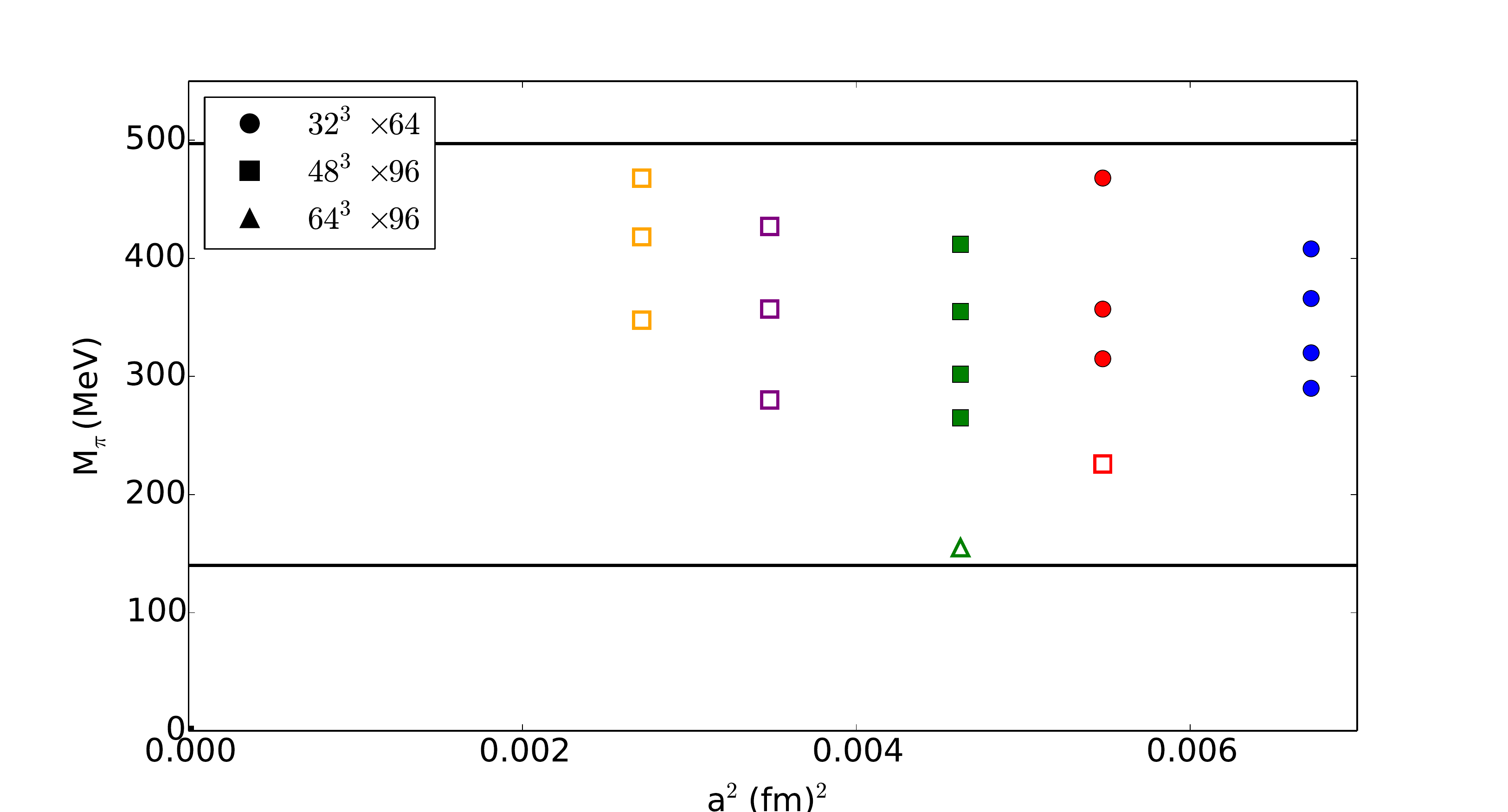}
  \caption{Lattice ensembles that will be used in this work characterised by pion mass, $m_\pi$, and lattice spacing, $a$. The black lines represent the physical pion and kaon masses, where the continuum limit occurs as $a\rightarrow0$. The solid points represent the ensembles used in this proceeding while the open points are still being finalised.}
  \label{fig:config}
\end{figure}
In our simulations, we have kept the bare quark mass, $\bar{m}=(m_u+m_d+m_s)/3$, held fixed at its physical value, while systematically varying the quark masses around the $SU(3)$ flavour symmetric point, $m_u=m_d=m_s$, to eventually extrapolate results at the physical point. We also have degenerate $u$ and $d$ quark masses, $m_u=m_d\equiv m_l$, restricting ourselves to $n_f=2+1$. The lattice spacings and pion masses are represented graphically in Fig.~\ref{fig:config}. Here the solid points represent the ensembles reported in this proceeding while the open points are still being finalised. We have five lattice spacings, $a= 0.082,~0.074,~0.068,~0.059,~0.0521$~fm enabling an extrapolation to the continuum limit as well as three lattice volumes $32^3\times 64$, $48^3\times 96$ and $64^3 \times 96$, which are indicated by the shapes of the points in Fig.~\ref{fig:config}.


\section{Calculating Matrix Elements using The Feynman-Hellmann Theorem}
The Feynman-Hellmann (FH) theorem is used to calculate hadronic matrix elements in lattice QCD through modifications to the QCD Lagrangian. Consider the following modification to the action  of our theory:
\begin{align}
S\rightarrow S+\lambda\mathcal{O}.
\label{eqn:modac}
\end{align}
Then the FH theorem as shown in Ref.~\cite{Horsley:2012pz, FH1}, provides a relationship between an energy shift and a matrix element of interest:
\begin{align}
\frac{\partial E_{X,\lambda}(\vec{k})}{\partial\lambda}\Big|_{\lambda=0}=\frac{1}{2E_{X}(\vec{k})}\bra{X,\vec{k}}\mathcal{O}\ket{X,\vec{k}}.
\label{eqn:FH5}
\end{align}
Importantly, the right-hand side is the standard matrix element of the operator $\mathcal{O}$ inserted on the hadron, $X$, in the absence of any background field. In lattice calculations, we modify the action in Eq.~\ref{eqn:modac}, then we examine the behaviour of hadron energies as the parameter $\lambda$ changes, and extract the above matrix element from the slope at $\lambda=0$.
\subsection{Application and Implementation}
In order to calculate the tensor charge, the extra term we add to the QCD action is:
\begin{align}
S\rightarrow S+\lambda\int d^4xi\bar{q}(x)\sigma^{\mu\nu}\gamma_5q(x),
\label{eqn:FH6}
\end{align}
where we will take the case of each quark flavour, $q$, separately. The tensor charge is related to the following nucleon matrix element:
\begin{align}
\bra{\vec{p},\vec{s}}\mathcal{T}^{\mu\nu}\ket{\vec{p},\vec{s}}=\frac{2}{m_N}(s^\mu p^\nu -s^\nu p^\mu)\delta q,
\label{eqn:ope}
\end{align}
where $\mathcal{T}^{\mu\nu}=i\bar{q}\sigma^{\mu\nu}\gamma_5q$. In our simulations, we have chosen $\mu=3$, $\nu=4$:
\begin{align}
\bra{\vec{p},\vec{s}}\mathcal{T}^{34}\ket{\vec{p},\vec{s}}=2\delta q s^3, 
\label{eqn:op1}
\end{align}
where  $s^3$ is the direction of nucleon polarisation along the $z$ axis, hence the FH theorem in Eq.~\ref{eqn:FH5} then gives:
\begin{align}
\frac{\partial E^+_\lambda}{\partial\lambda}\Big|_{\lambda=0}=\delta q, && \frac{\partial E^-_\lambda}{\partial\lambda}\Big|_{\lambda=0}=-\delta q,
\end{align}
where $E^{+/-}$ denotes the energy of the hadron with spin up/down in the $z$ direction in the presence of the tensor background field (Eq.~\ref{eqn:FH6}) with strength $\lambda$. The energy as a function of $\lambda$ is therefore given by:
\begin{align}
E^{\pm}(\lambda)=E(0)\pm\lambda\delta q +\mathcal{O}(\lambda^2).
\label{eqn:DELTAE}
\end{align}
We have related the change in energy of the hadron state to the transverse spin contribution from the quark flavour $q$. Alternatively, due to the combination of $\pm\lambda$, the spin-down state with positive $\lambda$ is equivalent to the energy shift of the spin-up state with negative $\lambda$. The nucleon isovector tensor charge is then given by the difference between the up and down quark contributions:
\begin{align}
g_T=\delta u-\delta d.
\end{align}   
\subsection{Results}
  \vspace{-20pt}
\begin{figure}[H]
  \centering
  \subfigure[]{\includegraphics[scale=0.23]{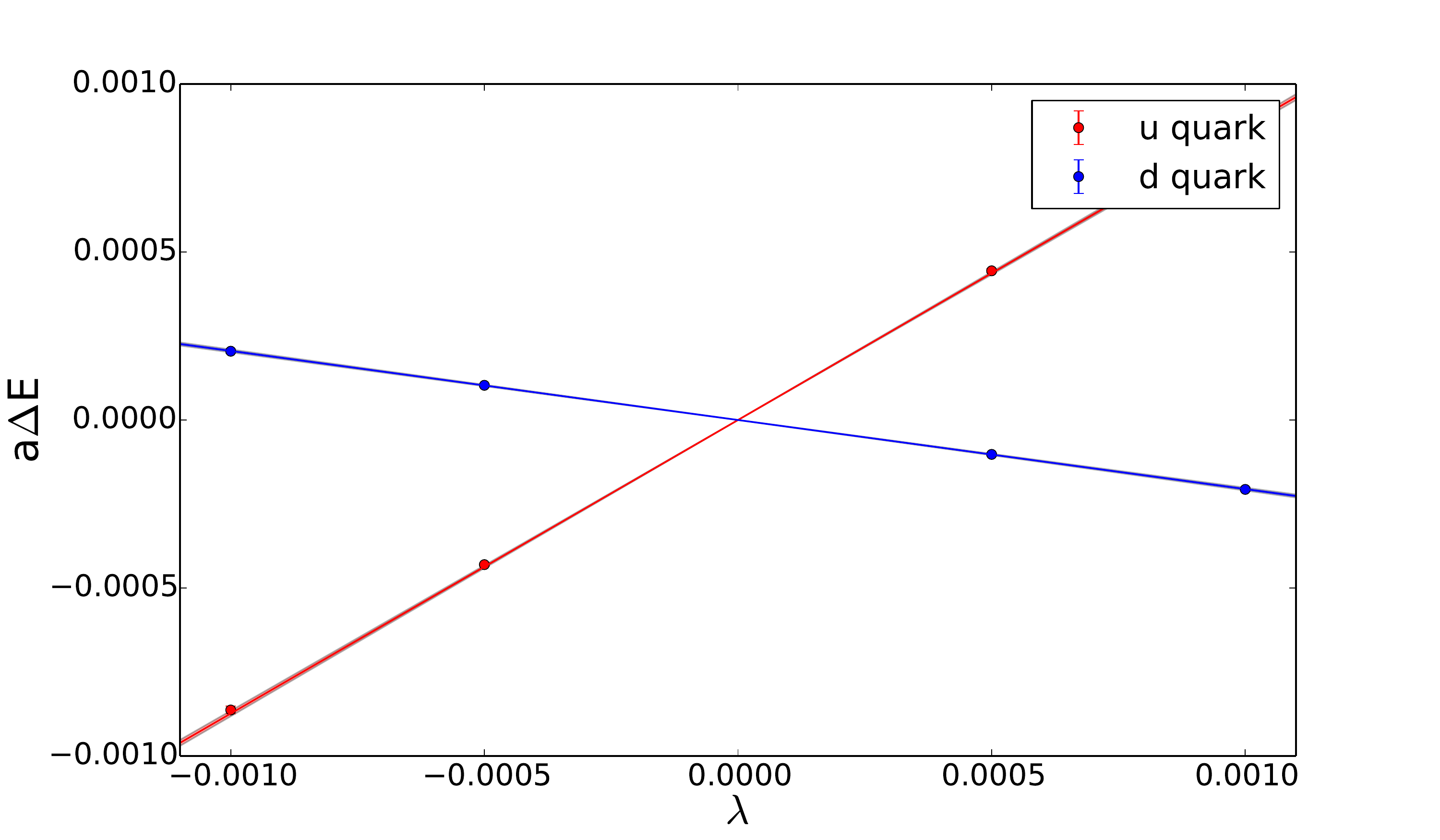}}
  \hskip -1ex
  \subfigure[]{\includegraphics[scale=0.23]{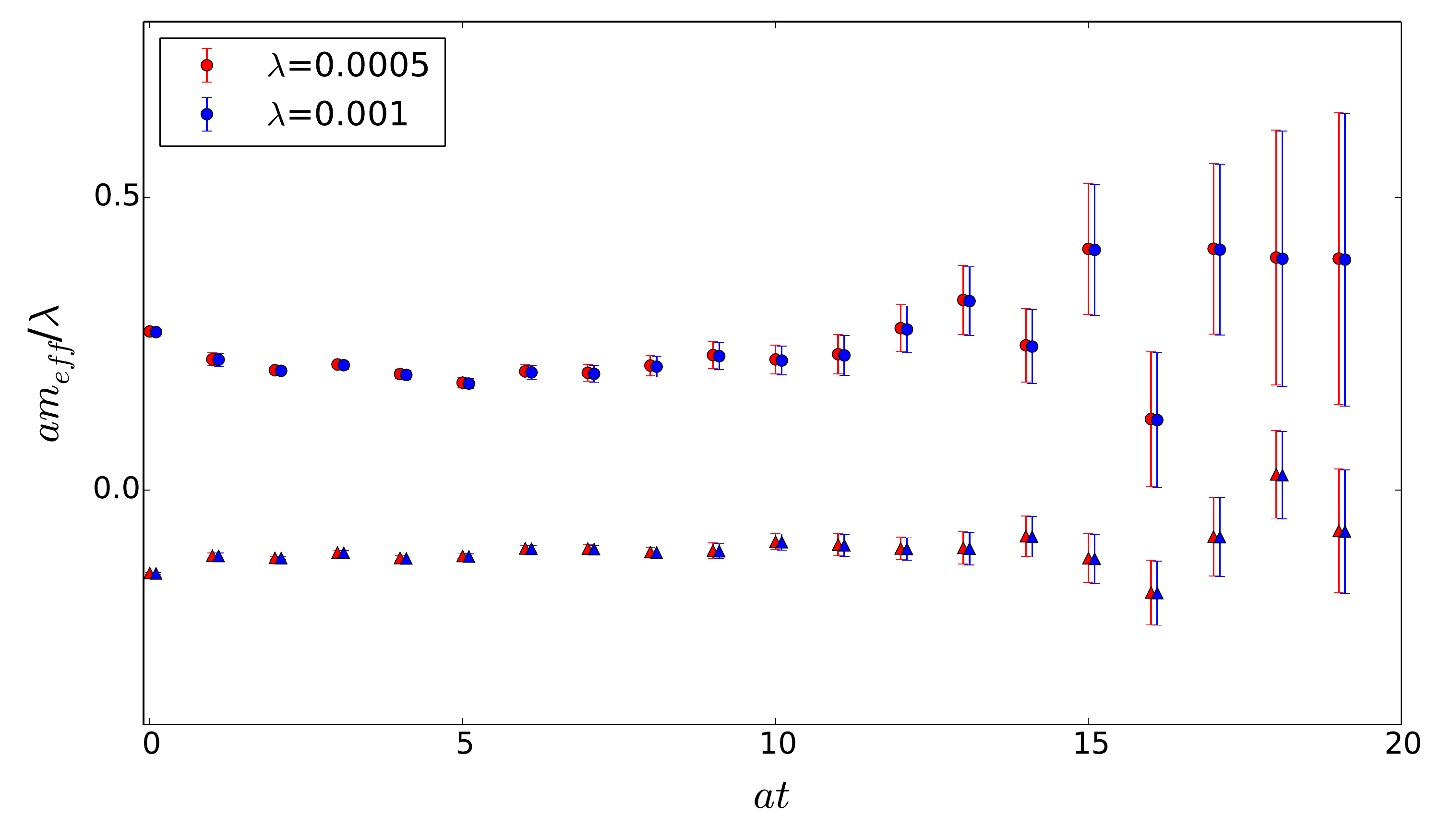}}
  \caption{\label{fig:FH1}(a) Change in proton energy for different parameter values, with a linear fit. Calculated at $a=0.082$fm, ($\kappa_l,~\kappa_s)=(0.119930,0.119930)$. Noting the error bars are smaller than the points displayed and we define $\Delta E=E_\lambda-E$. (b) Proton effective mass for the ratio (Eq.~\ref{eqn:corr_rat}) divided by $\lambda$, for the down quark at two different values of $\lambda$, for spin up (circle points) and spin-down (triangle points), calculated at $a=0.082$fm, ($\kappa_l,~\kappa_s)=(0.119930,0.119930)$. The points have been offset slightly for clarity.}
 \label{fig:FH1}
\end{figure}
We consider the ratio of two correlation functions, one calculated at $\lambda=0$ and the other at some finite value of $\lambda$. At sufficiently large Euclidean time, we isolate the energy difference by:
\begin{equation}
\begin{aligned}
  \frac{C_\lambda(t)}{C(t)}\stackrel{\text{large}~t}{=}&e^{-(E_\lambda-E)t}\frac{E}{E_\lambda}\frac{|A_\lambda|^2}{|A|^2}.
\end{aligned} 
\label{eqn:corr_rat}
\end{equation}
In Fig.~\ref{fig:FH1}(a) we plot the calculated nucleon energies for each value of $\lambda$. Note that while only two positive values of $\lambda$ are used, the energy shift of the spin-down state with positive $\lambda$ is equivalent to the energy shift of the spin-up state with negative $\lambda$. Negative $\lambda$ values hence come from flipping the spin of the nucleon. In Fig.~\ref{fig:FH1}(a) we perform a linear fit to Eq.~\ref{eqn:DELTAE} and by extracting the slope we get the following results:
\begin{align}
\delta u~=~&0.8266(77),\\
\delta d~=~&-0.1945(39),
\end{align} 
renormalised at $\mu=2\text{GeV}$ in the $\overline{\text{MS}}$ scheme \cite{Bikerton_2020, Constantinou:2014fka}. The Feynman-Hellmann theorem has some advantages over standard methods. Since hadron energies are extracted from two-point functions, control of excited state contamination in the Feynman-Hellmann is simplified compared to standard three-point analyses. Fig.~\ref{fig:FH1}(b) shows the effective mass for the ratio (Eq.~\ref{eqn:corr_rat}) divided by $\lambda$ for the down quark at two different values of $\lambda$, for spin-up and spin-down. The advantage of Feynman-Hellmann can be seen in Fig.~\ref{fig:FH1}(b), where we see a plateau in the effective mass and a control over excited state contamination.\\
The above process has been repeated for all pion masses on each of the lattice spacings as well as for the $\Sigma$ and $\Xi$ baryons.\\
Now that we have the quark contributions for multiple lattice ensembles, we use a $SU(3)$ flavour symmetry breaking method to extrapolate results for the tensor charge to the physical quark mass.

\section{Flavour Symmetry Breaking}
 As described above, here we keep the bare quark mass held fixed at approximately its physical value, while systematically varying the quark masses around the $SU(3)$ flavour symmetric point, to eventually extrapolate results to the physical point. \\

When $SU(3)$ is unbroken all octet baryon matrix elements of a given octet operator can be expressed in terms of just two couplings $f$ and $d$. However, once $SU(3$) is broken and we move away from the symmetric point we can construct quantities ($D_i$, $F_i$) which are equal at the symmetric point but differ in the case where the quark masses are different. The theory behind constructing these quantities is described in Ref.~\cite{Bickerton_2019}. The result of constructing these quantities leads to `fan' plots, with slope parameters ($r_i$, $s_i$) relating them. Following the method in Ref.~\cite{Bickerton_2019} we use the fan plots to extrapolate the tensor charge to the physical point.
\subsection{Mass Dependence: `Fan Plots'}
We hold the average quark masses, $\bar{m}$, fixed, while moving away from the symmetric point. Hence we only consider the non-singlet polynomials in the quark mass. In this section quantities $(D_i, F_i)$ are constructed  which are equal at the symmetric point and differ in the case where the quark masses are different, we then can evaluate the the violation of $SU(3)$ symmetry that emerges from the difference in $m_s-m_l$. Here we introduce the notation for the matrix element transition of $B\rightarrow B^\prime$ is as follows:
\begin{align}
A_{\bar{B'}FB}=\bra{B'}J^F\ket{B},
\end{align}
where $J^F$ is the appropriate operator from Ref.~\cite{Bickerton_2019}, and $F$ represents the flavour structure of the operator.
\subsection{The \textit{d}-fan} 
Following Ref.~\cite{Bickerton_2019}, we construct the following combinations of matrix elements:
\begin{equation}
\begin{aligned}
D_1\equiv-(A_{\bar{N}\eta N}+A_{\bar{\Xi}\eta \Xi})~=~&2d-2r_1\delta m_l,\\
D_2\equiv A_{\bar{\Sigma}\eta \Sigma}~=~&2d+(r_1+2\sqrt{3}r_3)\delta m_l,\\
D_4\equiv\frac{1}{\sqrt{3}}(A_{\bar{N}\pi N}-A_{\bar{\Xi}\pi \Xi})~=~&2d-\frac{4}{\sqrt{3}}r_3\delta m_l,\\
D_6\equiv \frac{1}{\sqrt{6}}(A_{\bar{N}K\Sigma}+A_{\bar{\Sigma}K\Xi})~=~&2d+\frac{2}{\sqrt{3}}r_3\delta m_l,\\
\label{eqn:Dfan}
\end{aligned}
\end{equation}
where $\delta m_l=m_l-\bar{m}$. The quantities $D_i$ can be calculated for each quark mass we calculated on the lattice. For example:
\begin{equation}
\begin{aligned}
D_1~=~&-(A_{\bar{N}\eta N}+A_{\bar{\Xi}\eta \Xi})\\
  =~&-\left(\frac{1}{\sqrt{6}}(\delta u_p+\delta d_{p})+\frac{1}{\sqrt{6}}(\delta u_\Xi-2\delta s_\Xi)\right),
\end{aligned}
\end{equation}
where we introduce the notation $\delta q_B$ to denote the quark, $q$, tensor charge in the baryon, $B$. Here $\delta u_p$, $\delta d_p$, $\delta u_\Xi$ and $\delta s_\Xi$ are the results calculated using the FH theorem for each lattice ensemble. An `average D' can also be constructed from the diagonal amplitudes:   
\begin{align}
X_D=\frac{1}{6}(D_1+2D_2+3D_4)=2d+\mathcal{O}(\delta m_l^2),
\label{eqn:avD}
\end{align}
 which is constant in $\delta m_l$ up to terms $\mathcal{O}(\delta m_l^2)$.
\subsection{The \textit{f}-fan}
Similarly another five quantities, $F_i$, can be constructed which all have the same value, $2f$, at the $SU(3)_f$ symmetric point:
\begin{equation}
\begin{aligned}
F_1\equiv \frac{1}{\sqrt{3}}(A_{\bar{N}\eta N}-A_{\bar{\Xi}\eta \Xi})~=~&2f-\frac{2}{\sqrt{3}}s_2\delta m_l,\\
F_2\equiv (A_{\bar{N}\pi N}+A_{\bar{\Xi}\pi \Xi}~=~&2f+4s_1\delta m_l,\\
F_3\equiv A_{\bar{\Sigma}\pi\Sigma}~=~&2f+(-2s_1+\sqrt{3}s_2)\delta m_l,\\
F_4\equiv\frac{1}{\sqrt{2}}(A_{\bar{\Sigma}K \Xi}-A_{\bar{N}K \Sigma})~=~&2f-2s_1\delta m_l,\\
F_5\equiv \frac{1}{\sqrt{3}}(A_{\bar{\Lambda}K \Xi}-A_{\bar{N}K \Lambda})~=~&2f+\frac{2}{\sqrt{3}}(\sqrt{3}s_1-s_2)\delta m_l.\\
\label{eqn:Ffan}
\end{aligned}
\end{equation}
Again, an `average F' can be calculated through:  
\begin{align}
X_F=\frac{1}{6}(3F_1+F_2+2F_3)=2f+\mathcal{O}(\delta m_l^2).
\label{eqn:avF}
\end{align}
In this work, only the connected quark-line terms are computed. Quark-line disconnected terms only show up in the $r_1$ coefficient and $r_1^{\text{discon}}$ cancels in the case $g_T=\delta u-\delta d$. Unlike the $d$-fan, the $f$-fan to linear order, has no error from dropping the quark-line disconnected contributions, as none of the $r_i$ parameters appear in the $f$-fan. 
\subsection{Results}
  \vspace{-20pt}
\begin{figure}[h]
  \centering
  \subfigure[]{\includegraphics[scale=0.27]{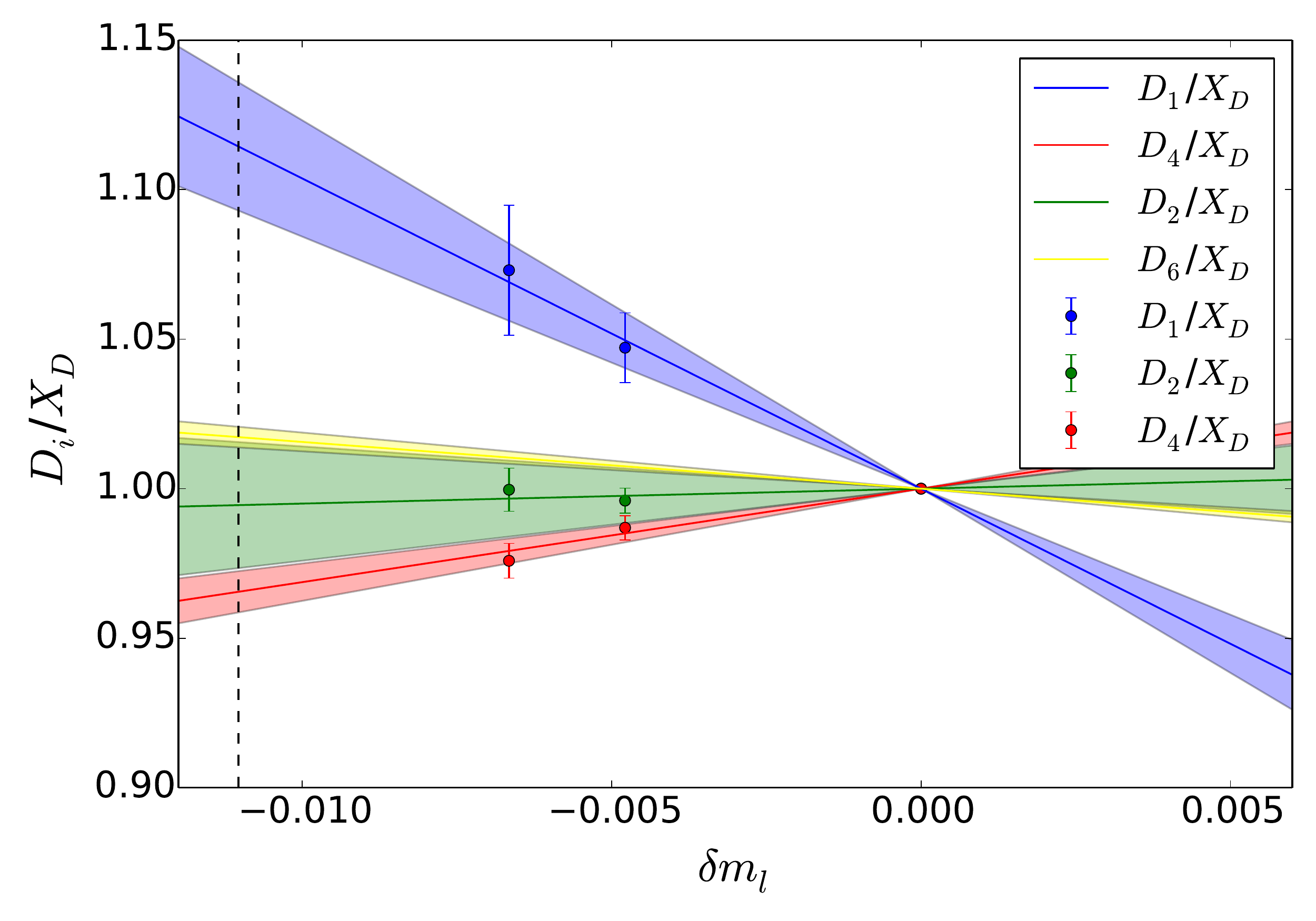}}
  \hskip -1ex
  \subfigure[]{\includegraphics[scale=0.27]{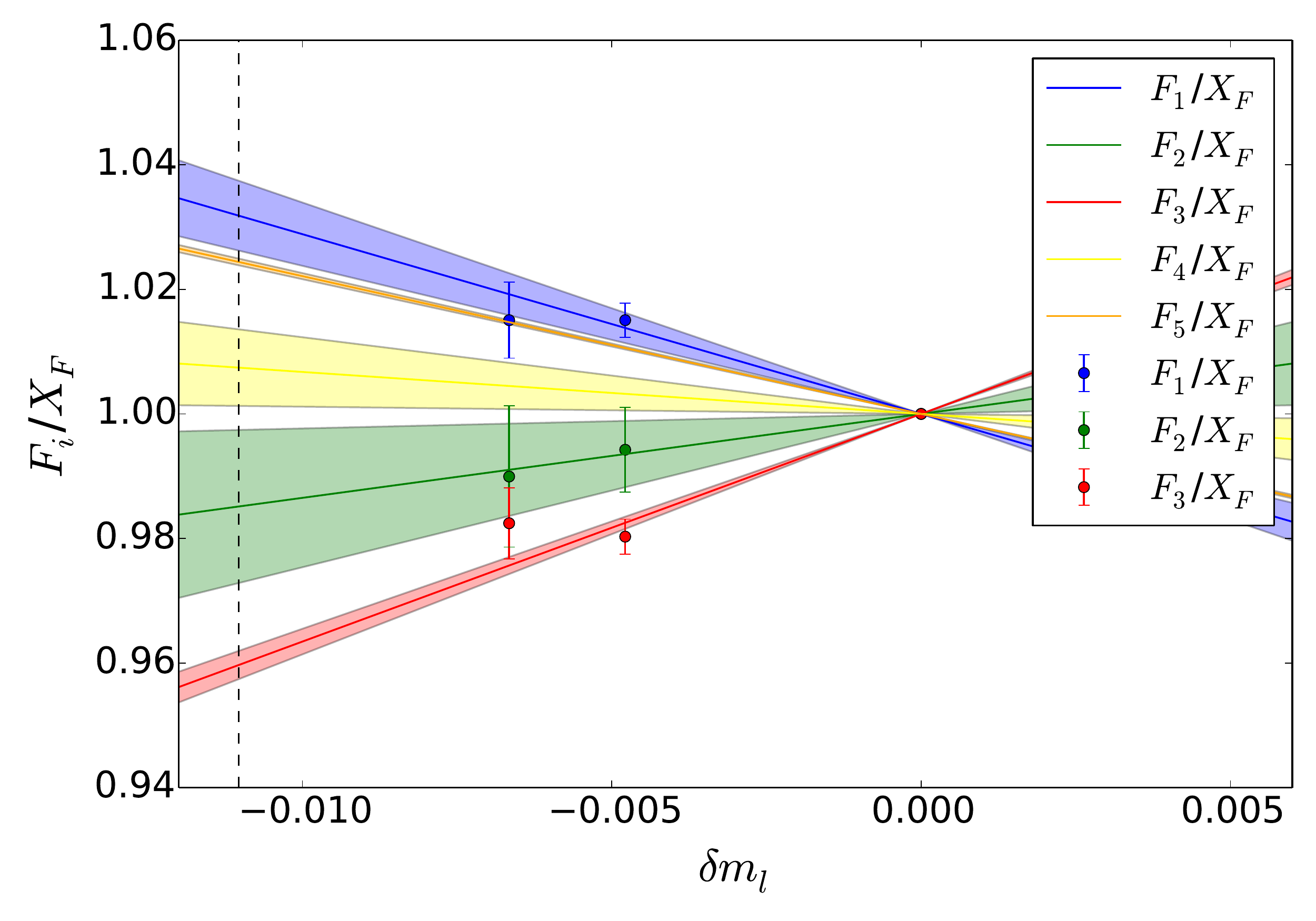}}
  \caption{\label{fig:fan}(a) The three fits $D_1$, $D_2$ and $D_4$ (b) The three fits $F_1$, $F_2$ and $F_3$. The vertical black dotted line represents the physical point. Results for the three ensembles with the heaviest pion masses at $a=0.074$fm ensemble.}

\end{figure}
Fig.~\ref{fig:fan}(a) shows the `fan' plot $\tilde{D}_i=D_i/X_D$ for $i=1,~2$ and $4$. Here the lines correspond to the linear fits of the $D_i$ using Eq.~\ref{eqn:Dfan}. From these linear fits the slope parameters $\tilde{r_1}=r_1/X_D$ and $\tilde{r_3}=r_3/X_D$ are determined. These parameters can also predict the off-diagonal term for $i=6$, which is also shown. Similarly in Fig.~\ref{fig:fan}(b) we have the `fan' plot $\tilde{F}_i=F_i/X_F$ for $i=1,~2$ and $3$, where the lines correspond to the linear fits using Eq.~\ref{eqn:Ffan}. Similarly, the parameters $\tilde{s_1}=s_1/X_F$ and $\tilde{s_2}=s_2/X_F$ are determined from the linear fits. Again, the corresponding off-diagonal terms for $i=4,5$ were also predicted and plotted.\\
By forming appropriate linear combinations, we reconstruct the matrix elements in a particular hadron: 
\begin{equation}
\begin{aligned}
\bra{p}\bar{u}\Gamma u\ket{p}~=&~2\sqrt{2}f+\Big(\sqrt{\frac{3}{2}}r_1-\sqrt{2}r_3+\sqrt{2}s_1-\sqrt{\frac{3}{2}}s_2\Big)\delta m_l,&\\
  \bra{p}\bar{d}\Gamma d\ket{p}~=&~\sqrt{2}(f-\sqrt{3}d)+\Big(\sqrt{\frac{3}{2}}r_1-\sqrt{2}r_3-\sqrt{2}s_1-\sqrt{\frac{3}{2}}s_2\Big)\delta m_l,&
\end{aligned}
\end{equation}
and hence the nucleon isovector tensor charge: 
\begin{equation}
\begin{aligned}
g_T=\bra{p}\bar{u}\Gamma u\ket{p}-\bra{p}\bar{d}\Gamma d\ket{p},
\end{aligned}
\end{equation}
for $\Gamma=i\sigma^{34}\gamma_5$. To obtain an extrapolation of $g_T$ to the physical point, we evaluate with  $\delta m_l \rightarrow\delta m_l^*$. The physical quark mass point, $\delta m_l^*$, has been determined in Ref.~\cite{PhysRevD.91.074512}. A similar procedure is followed to determine the nucleon scalar charge.\\
In this proceeding the above method was applied using the ensembles represented by solid points in Fig.~\ref{fig:config}. The result for each lattice spacing was averaged, giving:
\begin{align}
g_S~=~&0.95(03)(14),\\
g_T~=~&1.014(45)(80),
\end{align} 
renormalised at $\mu=2\text{GeV}$ in the $\overline{\text{MS}}$ scheme \cite{Bikerton_2020, Constantinou:2014fka}. The first error in brackets is the statistical error and the last error is systematic. As these results are preliminary we have taken the systematic error to be half the difference between the maximum and minimum value of $g_T$ and $g_S$. Noting that discretisation and volume effects have not yet been quantified. These results are comparable to those given in the FLAG Review \cite{FlavourLatticeAveragingGroup:2019iem}.
  \vspace{-5pt}
\section{Impact of Lattice Results on Phenomenology}
  \vspace{-15pt}
\begin{figure}[H]
  \centering
  \subfigure[]{\includegraphics[scale=0.4]{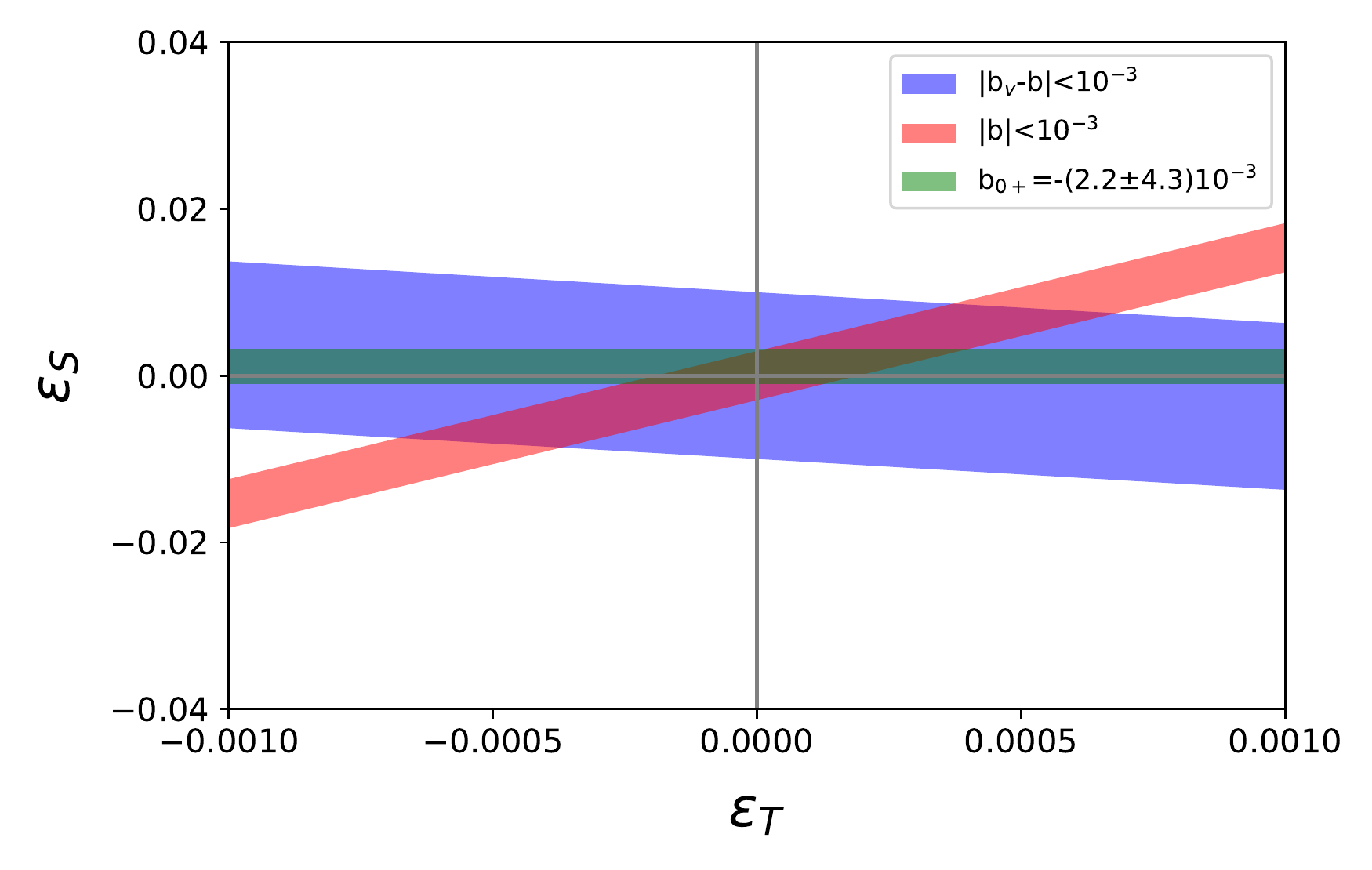}}
  \hskip -1ex
  \subfigure[]{\includegraphics[scale=0.4]{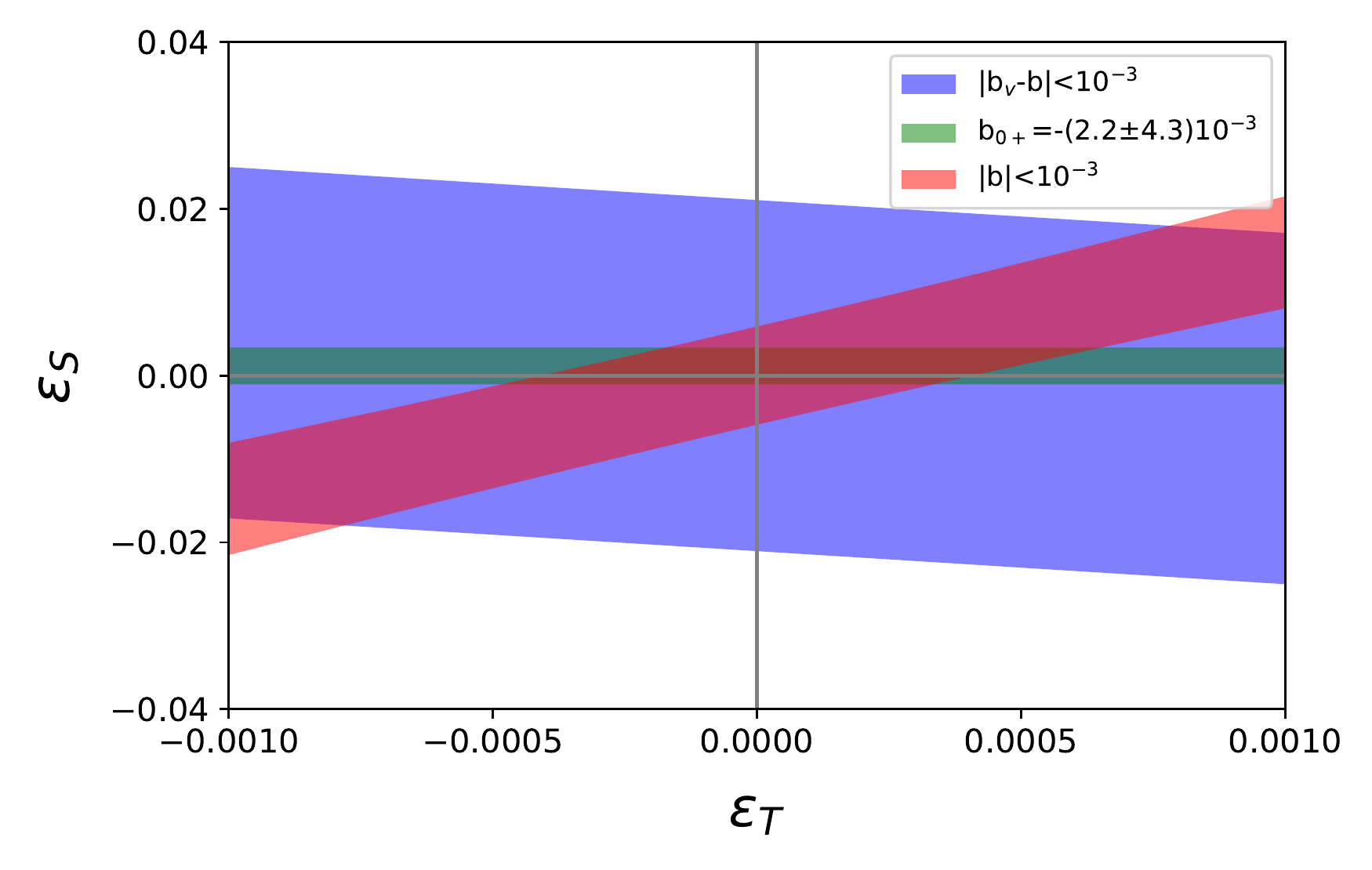}}
  \caption{\label{fig:bound}(a) Allowed regions in the $\epsilon_S-\epsilon_T$ plane for the scenario $g_S = g_T = 1$ with no uncertainty. The green band is the existing band on $b_{0+}$ \cite{Bhattacharya:2011qm, Hardy:2008gy}. (b) Tensor and scalar charges taken from lattice QCD with $g_S=0.95(03)(14)$ and $g_T=1.014(45)(80)$.}
\end{figure}

As the long term goal of this work is to support precision tests of the Standard Model, here we highlight the potential for improved precision from nucleon matrix elements in lattice QCD. Following the work of Ref.~\cite{Bhattacharya:2011qm}, in Fig.~\ref{fig:bound}(a) we show the constraint on the $\epsilon_S-\epsilon_T$ plane assuming perfect knowledge of the nucleon matrix elements. These current best constraints on scalar and tensor interactions arise from $0^+\rightarrow0^+$ nuclear beta decays and radioactive pion decay, which is shown by the green band~\cite{Bhattacharya:2011qm, Hardy:2008gy}. The neutron constraints are future projections at the $10^{-3}$ level, derived from Eq.~\ref{eqn:b} and Eq.~\ref{eqn:bv}, shown by the red and blue bands in Fig.~\ref{fig:bound}(a). For a more realistic constraint, including the hadronic uncertainties, in Fig.~~\ref{fig:bound}(b) we show the corresponding figure including our best-estimates from the preliminary results reported here for $g_S$ and $g_T$. When accounting for uncertainties in these lattice QCD calculations, the boundaries on the bands in Fig.~\ref{fig:bound}(b) become wider and contraining power is lost. In order to fully utilise the constraining power of future $10^{-3}$ experiments, understanding the lattice-QCD estimates of the tensor and scalar charge at the level of 10\% is required ~\cite{Bhattacharya:2011qm}.

\section{Conclusion}
In this work we have presented preliminary results for the nucleon tensor charge using the Feynman-Hellmann theorem, as well as using a flavour symmetry breaking method to systematically approach the physical quark mass. The Feynman-Hellmann theorem has advantages over using standard methods as control of excited state contamination is more simple than the standard three-point analyses. In the flavour symmetry breaking method we used, symmetry constraints  are automatically built in order-by-order in $SU(3)$ breaking. We have full coverage of $a$, $m_\pi$ and volume meaning in future we can control those systematics to reliably deliver desired precision goals in the future. 

\section*{Acknowledgements}
The numerical configuration generation (using the BQCD lattice QCD program \cite{Haar})) and data analysis (using the Chroma software library \cite{EDWARDS2005832}) was carried out on the DiRAC Blue Gene Q and Extreme Scaling (EPCC, Edinburgh, UK) and Data Intensive (Cambridge, UK) services, the GCS supercomputers JUQUEEN and JUWELS (NIC, Jülich, Germany) and resources provided by HLRN (The North-German Supercomputer Alliance), the NCI National Facility in Canberra, Australia (supported by the Australian Commonwealth Government) and the Phoenix HPC service (University of Adelaide). RH is supported by STFC through grant ST/P000630/1. HP is supported by DFG Grant No. PE 2792/2-1. PELR is supported in part by the STFC under contract ST/G00062X/1. GS is supported by DFG Grant No. SCHI 179/8-1. RDY and JMZ are supported by the Australian Research Council grant DP190100297.
\bibliographystyle{h-physrev}
\bibliography{thesis_bib.bib}

%

\end{document}